\documentclass[12pt]{article}
\usepackage{amsmath,amssymb,amsthm}
\usepackage{graphicx,psfrag,epsf}
\usepackage{enumerate}
\usepackage{natbib}
\usepackage{url} 
\usepackage{bm}
\usepackage[hidelinks]{hyperref}
\usepackage{float}

\newcommand{\blind}{1}

\def\ctimes{\times \cdots \times}
\def\Y{\mathbf{Y}}
\def\I{\mathbf{I}}
\def\U{\mathbf{U}}
\def\u{\mathbf{u}}
\def\X{\mathbf{X}}
\def\XX{\mathbb{X}}
\def\YY{\mathbb{Y}}
\def\E{\mathbf{E}}
\def\EE{\mathbb{E}}
\def\V{\mathbf{V}}
\def\A{\mathbf{A}}
\def\a{\mathbf{a}}
\def\B{\mathbf{B}}
\def\AA{\mathbb{A}}
\def\BB{\mathbb{B}}
\def\B{\mathbf{B}}
\def\b{\mathbf{b}}
\def\C{\mathbf{C}}
\def\CC{\mathbb{C}}
\def\d{\mathbf{d}}
\def\D{\mathbf{D}}
\def\v{\mathbf{v}}

\def\mmu{\boldsymbol{\mu}}
\def\SSigma{\boldsymbol{\Sigma}}
\def\log{\mbox{log}}

\def\new{\mbox{new}}
\def\tp#1{[\![#1]\!]}

\newcommand{\tprod}[3] {
  \langle #1, #2 \rangle_{#3}}
 \DeclareMathOperator*{\amin}{arg\,min}
\def\argmin#1{\underset{#1}{\amin}}
 \DeclareMathOperator*{\rank}{rank}
  \DeclareMathOperator*{\vect}{vec}
  \DeclareMathOperator*{\pr}{pr}
 \newtheorem{prop}{Proposition} 
\newtheorem{theorem}{Theorem}

\begin{document}

\def\spacingset#1{\renewcommand{\baselinestretch}%
{#1}\small\normalsize} \spacingset{1}


\if1\blind
{
  \title{\bf Tensor-on-tensor regression}
  \author{Eric F. Lock\thanks{
    The author gratefully acknowledges the support of NIH grant  ULI RR033183/KL2 RR0333182.}\hspace{.2cm}\\
    Division of Biostatistics, University of Minnesota}
  \maketitle
} \fi

\if0\blind
{
  \bigskip
  \bigskip
  \bigskip
  \begin{center}
    {\LARGE\bf Tensor-on-tensor regression}
\end{center}
  \medskip
} \fi

\bigskip
\begin{abstract}
We propose a framework for the linear prediction of a multi-way array (i.e., a tensor) from another multi-way array of arbitrary dimension, using the contracted tensor product.  This framework generalizes several existing approaches, including methods to predict a scalar outcome from a tensor, a matrix from a matrix, or a tensor from a scalar.  We describe an approach that exploits the multiway structure of both the predictors and the outcomes by restricting the coefficients to have reduced CP-rank.  We propose a general and efficient algorithm for penalized least-squares estimation, which allows for a ridge ($L_2$) penalty on the coefficients.  The objective is shown to give the mode of a Bayesian posterior, which motivates a Gibbs sampling algorithm for inference.  We illustrate the approach with an application to facial image data.  An R package is available at \url{https://github.com/lockEF/MultiwayRegression}. 
\end{abstract}

\noindent%
\vfill

\newpage
\spacingset{1.05} 
\section{Introduction}
\label{intro}

For many applications data are best represented in the form of a \emph{tensor}, also called a \emph{multi-way} or \emph{multi-dimensional} array, which extends the familiar two-way data matrix (\emph{Samples} $\times$ \emph{Variables}) to higher dimensions.   Tensors are increasingly encountered in fields that require the automated collection of high-throughput data with complex structure.  For example, in molecular ``omics"  profiling it is now common to collect high-dimensional data over multiple subjects, tissues, fluids or time points within a single study.  For neuroimaging modalities such as fMRI and EEG, data are commonly represented as multi-way arrays with dimensions that can represent subjects, time points, brain regions, or frequencies.  In this article we consider an application to a collection of facial images from the Faces in the Wild database \citep{learned2016labeled}, which when properly aligned to a $90 \times 90$ pixel grid can be represented as a 4-way array with dimension \emph{Faces} $\times$ \emph{X} $\times$ \emph{Y} $\times$ \emph{Colors}, where \emph{X} and \emph{Y} give the horizontal and vertical location of each pixel.     

This article concerns the prediction of an array of arbitrary dimension $Q_1 \ctimes Q_M$ from another array of arbitrary dimension $P_1 \ctimes P_L$.  For $N$ training observations, this involves an outcome array $\YY: N \times Q_1 \ctimes Q_M$ and a predictor array $\XX: N \times P_1 \ctimes P_L$.  For example, we consider the simultaneous prediction of several describable  attributes for faces from their images \citep{kumar2009attribute}, which requires predicting the array $\YY$: \emph{Faces} $\times$ \emph{Attributes} from $\XX$: \emph{Faces} $\times$ \emph{X} $\times$ \emph{Y} $\times$ \emph{Colors}.  Other potential applications include the prediction of fMRI from EEG data (see \citet{de2011predicting}) and the prediction of gene expression across multiple tissues from other genomic variables (see \cite{ramasamy2014genetic}).   

The task of tensor-on-tensor regression extends a growing literature on the predictive modeling of tensors under different scenarios.  Such methods commonly rely on tensor factorization techniques \citep{kolda2009tensor}, which reconstruct a tensor using a small number of underlying patterns in each dimension.  Tensor factorizations extend well known techniques for a matrix, such as the singular value decomposition and principal component analysis,  to higher-order arrays.   A classical and straightforward technique is the PARAFAC/CANDECOMP (CP) \citep{harshman1970foundations} decomposition, in which the data are approximated as a linear combination of rank-1 tensors.  An alternative is the Tucker  decomposition \citep{tucker1966some}, in which a tensor is factorized into basis vectors for each dimension that are combined using a smaller core tensor.  The CP factorization is a special case of the Tucker factorization wherein the core tensor is diagonal.  Such factorization techniques are useful to account for and exploit multi-way dependence and reduce dimensionality.  

 Several methods have been developed for the prediction of a scalar outcome from a tensor of arbitrary dimension: $\YY: N \times 1$ and $\XX: N \times P_1 \ctimes P_L$.  \citet{zhou2013tensor} and \citet{guo2012tensor} propose tensor regression models for a single outcome in which the coefficient array is assumed to have a low-rank CP factorization.  The proposed framework in \citet{zhou2013tensor} extends to generalized linear models and allows for the incorporation of sparsity-inducing regularization terms.  An analogous approach in which the coefficients are assumed to have a Tucker structure is described by \citet{li2013}. Several methods have also been developed for the classification of multiway data (categorical $Y: N \times 1$) \citep{tao2007supervised,wimalawarne2016,lyu2017discriminating}, extending well-known linear classification techniques under the assumption that model coefficients have a factorized structure. 
 
There is also a wide literature on the prediction of a matrix from another matrix, $\YY: N \times Q$ and $\XX: N \times P$. A classical approach is reduced rank regression, in which the $P \times Q$ coefficient matrix is restricted to have low rank \citep{izenman1975reduced,mukherjee2011reduced}.  \citet{miranda2015} describe a Bayesian formulation for regression models with multiple outcome variables and multiway predictors ($\YY: N \times Q$ and $\XX: N \times P_1 \ctimes P_L$), which is applied to a neuroimaging study.   Conversely, tensor response regression models have been developed to predict a multiway outcome from vector predictors ($\YY: N \times Q_1 \ctimes Q_M$, $\XX: N \times P$).    \citet{sun2016sparse} propose a tensor response regression wherein a multiway outcome is assumed to have a CP factorization, and  \citet{li2016parsimonious} propose a tensor response regression wherein a multiway outcome is assumed to have a Tucker factorization with weights determined by vector-valued predictors.  For a similar context \citet{lock2016supervised} describe a supervised CP factorization, wherein the components of a CP factorization are informed by vector-valued covariates.    \cite{hoff2015multilinear} extend a bilinear regression model for matrices to the prediction of an outcome tensor from a predictor tensor with the same number of modes (e.g., $\YY: N \times Q_1 \ctimes Q_K$ and $\XX: N \times P_1 \ctimes P_K$) via a Tucker product and describe a Gibbs sampling approach to inference.

The above methods address several important tasks, including scalar-on-tensor regression, vector-on-vector regression, vector-on-tensor regression and tensor-on-vector regression. However, there is a lack of methodology to addresses the important and increasingly relevant task of tensor-on-tensor regression, i.e., predicting an array of arbitrary dimension from another array of arbitrary dimension.  This scenario is considered within a comprehensive theoretical study of convex tensor regularizers \cite{raskutti2015convex}, including the tensor nuclear norm.  However, they do not discuss estimation algorithms for this context, and computing the tensor nuclear norm is  NP-hard \citep{sun2016sparse,friedland2014nuclear}. In this article we propose a \emph{contracted tensor product} for the linear prediction of a tensor $\XX$ from a tensor $\YY$, where both $\XX$ and $\YY$ have arbitrary dimension, through a coefficient array $\BB$ of dimension $P_1 \ctimes P_L \times Q_1 \ctimes Q_M$.  This framework is shown to accommodate all valid linear relations between the variates of $\XX$ and the variates of $\YY$.  In our implementation $\BB$ is assumed to have reduced CP-rank, a simple restriction which simultaneously exploits the multi-way structure of both $\XX$ and $\YY$ by borrowing information across the different modes and reducing dimensionality.  We propose a general and efficient algorithm for penalized least-squares estimation, which allows for a ridge ($L_2$) penalty on the coefficients.  The objective is shown to give the mode of a Bayesian posterior, which motivates a Gibbs sampling algorithm for inference.  

The primary novel contribution of this article is a framework and methodology that allows for tensor-on-tensor regression with arbitrary dimensions. Other novel contributions include optimization under a ridge penalty on the coefficients and Gibbs sampling for inference, and these contributions are also relevant to the more familiar special cases of tensor regression (scalar-on-tensor), reduced rank regression (vector-on-vector), and tensor response regression (tensor-on-vector).    

\section{Notation and Preliminaries}
\label{notation}

Throughout this article bold lowercase characters ($\mathbf{a}$) denote vectors, bold uppercase characters ($\mathbf{A}$) denote matrices, and uppercase blackboard bold characters ($\AA$) denote multi-way arrays of arbitrary dimension.  

Define a $K$-way array (i.e., a $K$th-order tensor) by $\AA: I_1 \ctimes I_K$, where $I_k$ is the dimension of the $k$th \emph{mode}.  The entries of the array are defined by indices enclosed in square brackets, $\AA[i_1,\hdots,i_K]$, where $i_k \in \{1,\hdots,I_k\}$ for $k \in 1,\hdots, K$.  

For vectors $\a_1,\hdots,\a_K$ of length $I_1,\hdots, I_K$, respectively, define the \emph{outer product}
\[
\AA = \a_1 \circ \a_2 \cdots \circ \a_K
\]
as the $K$-way array of dimensions $I_1 \ctimes I_K$, with entries
\[\AA[i_1,\hdots,i_K] = \prod_{k=1}^K \a_{k}[i_k]. \]
The outer product of vectors is defined to have \emph{rank} 1.
For matrices $\A_1, \hdots, \A_K$ of the same column dimension $R$, we introduce the notation
\begin{align}
\tp{\A_1,\hdots,\A_K} = \sum_{r=1}^R \a_{1r} \circ \cdots \circ \a_{Kr},	 \label{tpRank} 
\end{align}
where $\a_{kr}$ is the $r$th column of $\A_k$.  This gives a CP factorization, and an array that can be expressed in the form (\ref{tpRank}) is defined to have rank R.   

The vectorization operator $\vect(\cdot)$ transforms a multiway array to a vector containing the array entries.  Specifically, $\vect(\AA)$ is a vector of length $\prod_{k=1}^K I_K$ where 
\[\vect(\AA)\left[i_1+ \sum_{k=2}^K \left(\prod_{l=1}^{k-1} I_l \right) (i_k-1)\right] = \AA[i_1,\hdots,i_K].\] 
It is often useful to represent an array in matrix form via \emph{unfolding} it along a given mode.  For this purpose we let the rows of $\A^{(k)}: I_k \times \left(\prod_{j \neq k} I_j \right)$ give the vectorized versions of each \emph{subarray} in the $k$th mode.  

For two multiway arrays $\AA:I_1 \ctimes I_K \times P_1 \cdots P_L$ and $\BB:P_1 \ctimes P_L \times Q_1 \ctimes Q_M$ we define the \emph{contracted tensor product} 
\[\tprod{\AA}{\BB}{L}: I_1 \ctimes I_K \times Q_1 \ctimes Q_M\]
by 
\[\tprod{\AA}{\BB}{L}[i_1,\hdots i_K,q_1,\hdots,q_M] = \sum_{p_1=1}^{P_1} \cdots \sum_{p_L=1}^{P_L} \AA[i_1,\hdots i_K, p_1, \hdots, p_L] \BB[p_1,\hdots p_L, q_1, \hdots, q_M]. \]
An analogous definition of the contracted tensor product, with slight differences in notation, is given  in \citet{bader2006algorithm}.  Note that for matrices $\A: I \times P$ and $\B: P \times Q$,
\[\tprod{\A}{\B}{1} = \A \B,\]
and thus the contracted tensor product extends the usual matrix product to higher-order operands.  
    
\section{General framework}
\label{framework}

Consider predicting a multiway array $\YY: N \times Q_1 \ctimes Q_M$ from a multiway array $\XX: N \times P_1 \ctimes P_L$ with the model
\begin{align}
\YY = \tprod{\XX}{\BB}{L} + \EE 	\label{arrayEq}
\end{align}
where $\BB: P_1 \ctimes P_L \times Q_1 \ctimes Q_M$ is a coefficient array and $\EE: N \times Q_1 \ctimes Q_M$ is an error array.  The first $L$ modes of $\BB$ contract the dimensions of $\XX$ that are not in $\YY$, and the last $M$ modes of $\BB$ expand along the modes in $\YY$ that are not in $\XX$. The predicted outcome indexed by $(q_1,\hdots, q_M)$ is 
\begin{align}
\YY[n,q_1,\ldots,q_M] \approx \sum_{p_1}^{P_1} \cdots  \sum_{p_L}^{P_L} \XX[N,p_1,\hdots,p_L] \BB[p_1,\hdots,p_L,q_1,\hdots,q_M]  \label{outEq}  
\end{align} 
for observations $n=1,\hdots,N$.  In (\ref{arrayEq}) we forgo the use of an  intercept term for simplicity, and assume that $\XX$ and $\YY$ are each centered to have mean $0$ over all their values.      
 
Let $P$ be the total number of predictors for each observation,  $P= \prod_{l=1}^L P_L$, and $Q$ be the total number of outcomes for each observation, $Q= \prod_{m=1}^M Q_M$.   Equation (\ref{arrayEq}) can be reformulated by rearranging the entries of $\XX$, $\YY$, $\BB$ and $\EE$ into matrix form
\begin{align}
\Y^{(1)} = \X^{(1)} \B  + \E^{(1)} \label{matrixEq} 	 	
\end{align}
where $\Y^{(1)}: N \times Q$, $\X^{(1)}: N \times P$, and $\E^{(1)}: N \times Q$ are the arrays $\YY$, $\XX$ and $\EE$ unfolded along the first mode.  The columns of $\B: P \times Q$ vectorize the first $L$ modes of $\BB$ (collapsing $\XX$), and the rows of $\B$ vectorize the last $M$ modes of $\BB$ (expanding to $\YY$):
\[\B \left[p_1+ \sum_{l=2}^L \left(\prod_{i=1}^{l-1} P_l \right) (p_l-1), q_1+ \sum_{m=2}^M \left(\prod_{i=1}^{m-1} Q_m \right) (q_m-1)  \right] = \BB[p_1,\hdots,p_L,q_1,\hdots,q_M].\]
From its matrix form (\ref{matrixEq}) it is clear that the general framework (\ref{arrayEq}) supports all valid linear relations between the $P$ variates of $\XX$ and the $Q$ variates of $\YY$.   
      

\section{Estimation criteria}
\label{objective}

Consider choosing $\BB$ to minimize the sum of squared residuals  
\[||\YY- \tprod{\XX}{\BB}{L}||_F^2.\] 
The unrestricted solution for $\BB$ is given by separate OLS regressions for each of the $Q$ outcomes in $\YY$, each with design matrix $\X^{(1)}$; this is clear from (\ref{matrixEq}), where the columns of $\B$ are given by separate OLS regressions of $\X^{(1)}$ on each column of $\Y^{(1)}$. Therefore, the unrestricted solution is not well-defined if $Q > N$ or more generally if $\X^{(1)}$ is not of full column rank.  The unrestricted least squares solution may be undesirable even if it is well-defined, as it does not exploit the multi-way structure of $\XX$ or $\YY$, and requires fitting 
\begin{align}
\prod_{l=1}^L P_l \prod_{m=1}^M Q_m \label{fullDim}
\end{align}
unknown parameters.  
Alternatively, the multi-way nature of $\XX$ and $\YY$ suggests a low-rank solution for $\BB$. The rank $R$ solution can be represented as
\begin{align}
\BB = \tp{\U_1,\hdots,\U_L, \V_1,\hdots,\V_M}, \label{Bfac}
\end{align}
where $\U_l: P_l \times R$ for $l=1,\hdots,L$ and $\V_m: Q_m \times R$ for $m=1,\hdots,M$.  The dimension of this model is 
\begin{align}
R \left (P_1+\cdots+P_L+Q_1+\cdots+Q_M \right ), \label{redDim}
\end{align} 
which can be a several order reduction from the unconstrained dimensionality (\ref{fullDim}). Moreover, the reduced rank solution allows for borrowing of information across the different dimensions of both $\XX$ and $\YY$.  However, the resulting least-squares solution
\begin{align}
\argmin{\rank(\BB)\leq R} ||\YY- \tprod{\XX}{\BB}{L}||_F^2. \label{lsEq}
\end{align}   
is still prone to over-fitting and instability if the model dimension (\ref{redDim}) is high relative to the number of observed outcomes, or if the predictors $\XX$ have multicollinearity that is not addressed by the reduced rank assumption (e.g., multicollinearity within a mode).  High-dimensionality and multicollinearity are both commonly encountered in application areas that involve multi-way data, such as imaging and genomics.  To address these issues we incorporate an $L_2$ penalty on the coefficient array $\BB$, 
\begin{align}
\argmin{\rank(\BB)\leq R} ||\YY- \tprod{\XX}{\BB}{L}||_F^2. + \lambda ||\BB||_F^2, \label{ridgeEq}
\end{align}
where $\lambda$ controls the degree of penalization.  This objective is equivalent to that of ridge regression when predicting a vector outcome $\YY: N \times 1$ from a matrix $\XX: N \times P$, where necessarily $R=1$.  
   
\section{Identifiability}
\label{ident}  

The general predictive model (\ref{arrayEq}) is identifiable for $\BB$, in that $\BB \neq \BB^*$ implies 
\[\tprod{\tilde{\XX}}{\BB}{L} \neq \tprod{\tilde{\XX}}{\BB^*}{L} \]  
for some $\tilde{\XX} \in \mathbb{R}^{P_1 \ctimes P_L}$.  To show this, note that if $\tilde{\XX}$ is an array with $1$ in position $[p_1,\hdots,p_L]$ and zeros elsewhere, then 
\[\tprod{\tilde{\XX}}{\BB}{L}[q_1,\hdots,q_M] = \BB [p_1,\hdots,p_M,q_1,\hdots,q_M].\]

However, the resulting components $\U_1,\hdots,\U_L,\V_1,\hdots,\V_M$ in the factorized representation of $\BB$ (\ref{Bfac}) are not readily identified.  Conditions for their identifiability are equivalent to conditions for the identifiability of the CP factorization, for which there is an extensive literature.  To account for arbitrary scaling and ordering of the components, we impose the restrictions
\begin{enumerate}
\item $||\u_{1r}||=\cdots=||\u_{Lr}||=||\v_{1r}||=\cdots=||\v_{Mr}||$ for $r=1\hdots,R$, and 
\item $||\u_{11}|| \geq ||\u_{12}|| \geq \cdots \geq ||\u_{1R}||$.
\end{enumerate}
The above restrictions are generally enough to ensure identifiability when $L+M \geq 3$ under verifiable conditions \citep{sidiropoulos2000uniqueness}.  If $L+M = 2$ (i.e., when predicting a matrix from a matrix, a 3-way array from a vector, or a vector from a 3-way array), then $\BB$ is a matrix and we require additional orthogonality restrictions:
\begin{enumerate}
\setcounter{enumi}{2}
\item $\u_{1r}^T \u_{1r^*} = 0$ for all $r \neq r^*$, or $\v_{1r}^T \v_{1r^*} = 0$ for all $r \neq r^*$.
\end{enumerate}
In practice these restrictions can be imposed post-hoc, after the estimation procedure detailed in Section~\ref{opt}.  For $L+M \geq 3$, restrictions (a) and (b) can be imposed via a re-ordering and re-scaling of the components.  For $L+M=2$, components that satisfy restrictions (a), (b) and (c) can be identified via a singular value decomposition of $\BB$.

\section{Special cases}
\label{relations}

Here we describe other methods that fall within the family given by the reduced rank ridge objective (\ref{ridgeEq}).  When predicting a vector from a matrix ($Q=0$, $P=1$), this framework is equivalent to standard ridge regression \citep{hoerl1970ridge},  which is equivalent to OLS when $\lambda=0$. Moreover, a connection between standard ridge regression and continuum regression \citep{sundberg1993continuum} implies that the coefficients obtained through ridge regression are proportional to partial least squares regression for some $\lambda=\lambda^*$, and the coefficients are proportional to the principal components of $\XX$ when $\lambda \rightarrow \infty$.  

When predicting a matrix from another matrix ($Q=1, P=1$), the objective given by  (\ref{ridgeEq}) is equivalent to reduced rank regression \citep{izenman1975reduced} when $\lambda=0$.  For arbitrary $\lambda$ the objective is equivalent to a recently proposed reduced rank ridge regression \citep{mukherjee2011reduced}. 

When predicting a scalar from a tensor of arbitrary dimension ($Q=0$, arbitrary $P$),   (\ref{ridgeEq}) is equivalent to tensor ridge regression \citep{guo2012tensor}.  \citet{guo2012tensor} use an alternating approach to estimation but claim that the subproblem for estimation of each $\U_l$ cannot be computed in closed form and resort to gradient style methods instead. On the contrary, our optimization approach detailed in Section~\ref{opt} does give a closed form solution to this subproblem (\ref{upU}). Alternatively, \citet{guo2012tensor} suggest the separable form of the objective, 
\begin{align}
\argmin{\rank(\BB)\leq R} ||\mathbf{y}- \tprod{\XX}{\BB}{L}||_F^2. + \lambda \sum_{l=1}^L||\U_l||_F^2. \label{ridgeEqSep}
\end{align}
This separable objective is also used by \citet{zhou2013tensor}, who consider a power family of penalty functions for predicting a vector from a tensor using a generalized linear model; their objective for a Gaussian response under $L_2$ penalization is equivalent to (\ref{ridgeEqSep}). The solution of the separable $L_2$ penalty depends on arbitrary scaling and orthogonality restrictions for identifiability of the $\U_l$'s.  For example, the separable penalty (\ref{ridgeEqSep}) is equivalent to the non-separable $L_2$ penalty (\ref{ridgeEq}) if the columns of $\U_2,\hdots,\U_L$ are restricted to be orthonormal.  

Without scale restrictions on the columns of $\U_l$, the solution to the separable $L_2$ penalty is equal to the solution for the non-separable penalty $||\BB||_{*}$ for $L=2$, where $||\BB||_{*}$ defines the nuclear norm (i.e., the sum of the singular values of $\BB$).  This interesting result is given explicitly in Proposition~\ref{prop1}, and its proof is given in Appendix~\ref{propproof}. 
\begin{prop}
\label{prop1}
For $\BB = \tp{\U_1,\U_2} = \U_1 \U_2^T$, where the columns of $\U_1$ and $\U_2$ are orthogonal,   
	\[\argmin{\rank(\BB)\leq R} ||\mathbf{y}- \tprod{\XX}{\BB}{2}||_F^2 + \lambda \sum_{l=1}^2||\U_l||_F^2 = \argmin{\rank(\BB)\leq R} ||\mathbf{y}- \tprod{\XX}{\BB}{2}||_F^2 + 2 \lambda ||\BB||_*.\]
\end{prop}
   
\section{Optimization}
\label{opt}
We describe an iterative procedure to estimate $\BB$ that alternatingly solves the objective (\ref{ridgeEq}) for the component vectors in each mode, $\{\U_1,\hdots,\U_L, \V_1,\hdots,\V_M\}$, with the others fixed.  

\subsection{Least-squares}
\label{lsAlg}
Here we consider the case without ridge regularization, $\lambda=0$, wherein the component vectors in each mode are updated via separate OLS regressions.  

To simplify notation we describe the procedure to update $\U_1$ with $\{\U_2,\hdots,\U_L,$ $\V_1,\hdots,\V_M\}$ fixed.  The procedure to update each of $\U_2,\hdots, \U_L$ is analogous, because the loss function is invariant under permutation of the $L$ modes of $\XX$.  

Define $\CC_r: N \times P_1 \times Q_1 \ctimes Q_M$ to be the contracted tensor product of $\XX$ and the $r$'th component of the CP factorization without $\U_1$:
\[\CC_r = \tprod{\XX}{\u_{2r} \circ \cdots \circ \u_{Lr} \circ \v_{1r} \circ \cdots \circ \v_{Mr}}{L-1}.\]   
Unfolding $\CC_r$ along the dimension corresponding to $P_1$ gives the design matrix to predict $\vect(Y)$ for the $r$'th column of $\U_1$, $\C_r: N Q \times P_1$.  Thus, concatenating these matrices to define $\C: N Q \times R P_1$ by $\C = \left[\C_1 \hdots  \C_R\right]$ gives the design matrix for all of the entries of $\U_1$,  which are updated via OLS: 
\begin{align}
\vect(\U_1) = \left(\C^T\C \right)^{-1}\C^T \vect(\YY). \label{upU}	
\end{align}
  
For the outcome modes we describe the procedure to update $\V_M$ with $\{\U_1,\hdots,\U_L,$ $\V_1,\hdots,\V_{M-1}\}$ fixed.  The procedure to update each of $\V_1,\hdots, \V_{L-1}$ is analogous, because the loss function is invariant under permutation of the $M$ modes of $\YY$. 

Let $\Y_M: Q_M \times N \prod_{m=1}^{M-1}Q_m$ be $\YY$ unfolded along the mode corresponding to $Q_M$.  Define $\D: N \prod_{m=1}^{M-1} Q_m  \times R$ so that the $r$'th column of $D$, $\d_r$, gives the entries of the contracted tensor product of $\XX$ and the $r$'th component of the CP factorization without $\V_M$:
\[\d_r = \vect \left( \tprod{\XX}{\u_{1r} \circ \cdots \circ \u_{Lr} \circ \v_{1r} \circ \cdots \circ \v_{(M-1) r}}{L} \right).\]   
The entries of $\V_{M}$ are then updated via $Q_M$ separate OLS regressions: 
\begin{align}
\V_M = (\D^T \D)^{-1} \D^T \Y_M^T.	\label{upV}	
\end{align}

\subsection{Ridge-regularization}
\label{ridgeAlg}
For $\lambda>0$, note that the objective $(\ref{ridgeEq})$ can be equivalently represented as an unregularized least squares problem with modified predictor and outcome arrays $\tilde{\XX}$ and $\tilde{\YY}$:
\begin{align*}
\argmin{\rank(\BB)\leq R} ||\tilde{\YY}- \tprod{\tilde{\XX}}{\BB}{L}||_F^2. \label{lsEqMod}	
\end{align*}
Here $\tilde{\XX}: \left(N + P \right) \times P_1 \ctimes P_L$ is the concatenation of $\XX$ and a tensor wherein each $P_1 \ctimes P_L$ dimensional slice has $\sqrt{\lambda}$ for a single entry and zeros elsewhere; $\tilde{\YY}: \left(N + P \right) \times Q_1 \ctimes Q_M$ is the concatenation of $\YY$ and a $P \times Q_1 \ctimes Q_M$ tensor of zeros.  Unfolding $\tilde{\XX}$ and $\tilde{\YY}$ along the first dimension yields the matrices 

\[\tilde{\X}^{(1)} =  \left[ \begin{array}{c}
\X^{(1)} \\
\sqrt{\lambda} \mathbf{I}_{P \times P} 
\end{array} \right] \, \, \, \text{ and } \, \, \, 
\tilde{\Y}^{(1)} =  \left[ \begin{array}{c}
\Y^{(1)} \\
\mathbf{0}_{P \times \prod_{m=1}^L Q_m} 
\end{array} \right],\]
where $\mathbf{I}$ is the identity matrix and $\mathbf{0}$ is a matrix of zeros.  

Thus, one can optimize the objective (\ref{ridgeEq}) via alternating least squares by replacing $\tilde{\XX}$ for $\XX$  and $\tilde{\YY}$ for $\YY$ in the least-squares algorithm of Section~\ref{lsAlg}.  However, $\tilde{\XX}$  and $\tilde{\YY}$ can be very large for high-dimensional $\XX$. Thankfully, straightforward tensor algebra shows that this is equivalent to a direct application of algorithm in Section~\ref{lsAlg} to the original data $\XX$ and $\YY$, with computationally efficient modifications to the OLS updating steps (\ref{upU}) and (\ref{upV}).  The updating step for $\U_{1}$ (\ref{upU}) is 
\begin{align}
\vect(\U_1) = \left(\C^T\C + \lambda (\U_2^T \U_2 \cdot \cdots \cdot \U_L^T \U_L \cdot \V_1^T \V_1 \cdot \cdots \cdot  \V_M^T \V_M) \otimes \mathbf{I}_{P_1 \times P_1}    \right)^{-1}\C^T \vect(\YY) \label{upU2}	
\end{align}
where $\cdot$ defines the dot product and $\otimes$ defines the Kronecker product.  The updating step for $\V_{M}$ (\ref{upV}) is 
\begin{align}
\V_M = (\D^T \D + \lambda (\U_1^T\U_1 \cdot \cdots \cdot \U_L^T\U_L \cdot \V_1^T \V_1 \cdot \cdots \cdot \V_{M-1}^T \V_{M-1}))^{-1} \D^T \Y_M^T.	\label{upV2}
\end{align}
 This iterative procedure is guaranteed to improve the regularized least squares objective (\ref{ridgeEq}) at each sub-step,  until convergence to a (potentially local) optimum.  Higher levels of regularization ($\lambda \rightarrow \infty$) tend to convexify the objective and facilitate convergence to a global optimum; a similar phenomenon is observed in \citet{zhou2013tensor}.  In practice we find that robustness to initial values and local optima is improved by a tempered regularization, starting with larger values of $\lambda$ that gradually decrease to the desired level of regularization.  

\subsection{Tuning parameter selection}
\label{paramSelect}

Selection of $\lambda$ and $R$ (if unknown) can be accomplished by assessing predictive accuracy with a training and test set, as illustrated in Section~\ref{app}. More generally, these parameters can be selected via K-fold cross-validation. This approach has the advantage of being free of model assumptions.    Alternatively, it is straightforward to compute the deviance information criterion (DIC) \citep{spiegelhalter2014deviance} for posterior draws under the Bayesian inference framework of Section~\ref{inf} and use this as a model-based heuristic to select both $\lambda$ and $R$.              

\section{Inference}
\label{inf}

In the previous sections we have considered optimizing a given criteria for point estimation, without specifying a distributional form for the data or even a philosophical framework for inference.  Indeed, the estimator given by the objective (\ref{ridgeEq}) is consistent under a wide variety of distributional assumptions, including those that allow for correlated responses or predictors. See Appendix~\ref{consistency} for more details on its consistency.   

For inference and uncertainty quantification for this point estimate, we propose a Markov chain Monte Carlo (MCMC) simulation approach.  This approach is theoretically motivated by the observation that (\ref{ridgeEq}) gives the mode of a Bayesian probability distribution.  There are several other reasons to use MCMC simulation for inference in this context, rather than (e.g.,) asymptotic normality of the global optimizer under an assumed likelihood model (see \citet{zhou2013tensor} and \citet{zhang2014tensor} for related results).  The algorithm in Section~\ref{opt} may converge to a local minimum, which can still be used as a starting value for MCMC.   Moreover, our approach gives a framework for full posterior inference on $\hat{\beta}$ over its rank $R$ support, the conditional mean for observed responses, and the predictive distribution for the response array given new realizations of the predictor array without requiring the identifiability of $\theta=\{\U_1, \hdots, \U_L, \V_1,\hdots, \V_M\}$. Inference for $\theta$ is also possible under the conditions of Section~\ref{ident}.         

If the errors $\EE$ have independent N$(0,\sigma^2)$ entries, the log-likelihood of $\YY$ implied by the general model (\ref{arrayEq}) is 
\[ \log \, \pr(\YY \mid \sigma^2, \BB, \XX) = \mbox{constant}-\frac{1}{2 \sigma^2} ||\YY- \tprod{\XX}{\BB}{L}||_F^2, \]
and thus the unregularized objective (\ref{lsEq}) gives the maximum likelihood estimate under the restriction rank$(\BB) = R$.  For $\lambda>0$, consider a prior distribution for $\BB$ that is proportional to the spherical Gaussian distribution with variance $\sigma^2/\lambda$ over the support of rank $R$ tensors:
\begin{align}
 \pr(\BB) \propto \begin{cases} \exp \left( -\frac{\lambda}{2\sigma^2} ||\BB||_F^2 \right) \text{ if } \rank(\BB) \leq R. \\ 0 \text{ otherwise,} \end{cases} \label{Bprior}
\end{align}  
The log posterior distribution for $\BB$ is
\begin{align} \log \, \pr(\BB \mid \YY, \XX, \sigma^2) = \mbox{constant}-\frac{1}{2 \sigma^2} \left( ||\YY- \tprod{\XX}{\BB}{L}||_F^2 + \lambda ||\BB||_F^2\right) 
\label{Bpost}	
\end{align}
where rank$(\BB)=R$, which is maximized by (\ref{ridgeEq}).   

Under the factorized form (\ref{Bfac}) the full conditional distributions implied by (\ref{Bpost}) for each of $\U_1,\hdots,\U_L, \V_1,\hdots,\V_M$ are multivariate normal.  For example, the full conditional for $\U_1$ is 
\[\pr(\vect(\U_1) \mid \U_2, \hdots,\U_L, \V_1,\hdots,\V_M, \YY, \XX, \sigma^2) = N(\mmu_1, \SSigma_1),\]
where $\mmu_1$ is the right hand side of (\ref{upU2}) and 
\[\SSigma_1 = \sigma^2 \left(\C^T\C + \lambda (\U_2^T \U_2 \cdot \cdots \cdot \U_L^T \U_L \cdot \V_1^T \V_1 \cdot \cdots \cdot  \V_M^T \V_M) \otimes \mathbf{I}_{P_1 \times P_1}    \right)^{-1}\]
where $\C$ is defined as in Section~\ref{lsAlg}.  The full conditional for $\V_M$ is 
\[\pr(\vect(\V_M) \mid \U_2, \hdots,\U_L, \V_1,\hdots,\V_M, \YY, \XX, \sigma^2) = N(\mmu_{L+M}, \SSigma_{L+M}),\]
where $\mmu_{L+M}$ is given by the right hand side of (\ref{upV2}) and 
\[\SSigma_{L+M} = \sigma^2 (\D^T \D + \lambda (\U_1^T\U_1 \cdot \cdots \cdot \U_L^T\U_L \cdot \V_1^T \V_1 \cdot \cdots \cdot \V_{M-1}^T \V_{M-1}))^{-1} \otimes \I_{Q_M \times Q_M}\]
The derivations of the conditional means and variances $\{\mmu_i,\SSigma_i\}_{i=1}^{L+M}$ are given in Appendix~\ref{appPost}.  When $\lambda=0$ the full conditionals correspond to a flat prior on $\BB$, $\pr(\BB) \propto 1$ for rank$(\BB)$ = R, and the posterior mode is given by the unregularized objective (\ref{lsEq}).

In practice we use a flexible Jeffrey's prior for $\sigma^2$, $\pr(\sigma^2) \propto 1/ \sigma^2$, which leads to an inverse-gamma (IG) full conditional distribution, 
\begin{align}
\pr(\sigma^2 \mid \BB, \YY, \XX) = \mbox{IG}\left(\frac{NQ}{2} , \frac{1}{2} ||\YY- \tprod{\XX}{\BB}{L}||_F^2\right). \label{postSig2}	
\end{align}

We simulate dependent samples from the marginal posterior distribution of $\BB$ by Gibbs sampling from the full conditionals of $\U_1,\hdots,\U_L, \V_1,\hdots,\V_M,$ and $\sigma^2$:
\begin{enumerate}
\item Initialize $\BB^{(0)}$ by the posterior mode (\ref{ridgeEq}) using the procedure in Section~\ref{opt}. \\

For samples $t=1,\hdots,T$, repeat (b) and (c):\\
\item Draw $\sigma^{2 (t)}$ from $P \left(\sigma^2 \mid \BB^{(t-1)}, \YY, \XX \right)$ as in (\ref{postSig2}).   
\item Draw $\BB^{(t)} = \tp{\U_1^{(t)},\hdots,\U_L^{(t)}, \V_1^{(t)},\hdots,\V_M^{(t)}}$, as follows:  
\begin{align*}
\U_1^{(t)} &\sim P \left(\U_1 \mid \U_2^{(t-1)}, \hdots,\U_L^{(t-1)}, \V_1^{(t-1)},\hdots,\V_M^{(t-1)}, \YY, \XX, \sigma^{2 (t)} \right)\\
& \vdots  \\ 
\U_L^{(t)} &\sim P \left(\U_L \mid \U_1^{(t)}, \hdots,\U_{L-1}^{(t-1)}, \V_1^{(t-1)},\hdots,\V_M^{(t-1)}, \YY, \XX, \sigma^{2 (t)} \right)  \\
\V_1^{(t)} &\sim P \left(\V_1 \mid \U_1^{(t)}, \hdots,\U_L^{(t)}, \V_2^{(t-1)},\hdots,\V_M^{(t-1)}, \YY, \XX, \sigma^{2 (t)} \right) \\
&  \vdots  \\
\V_L^{(t)} &\sim P \left(\V_M \mid \U_1^{(t)}, \hdots,\U_L^{(t)}, \V_1^{(t)},\hdots,\V_{M-1}^{(t)}, \YY, \XX, \sigma^{2 (t)} \right) .	
\end{align*}
 \end{enumerate}  
For the above algorithm $\U_1,\hdots,\U_L, \V_1,\hdots,\V_M$ serve as a parameter augmentation to facilitate sampling for $\BB$.  Interpreting the marginal distribution of each of the $\U_l's$ or $\V_m's$ separately requires careful consideration of their identifiability (see Section~\ref{ident}).  One approach is to perform a post-hoc transformation of the components at each sampling iteration \[\BB^{(t)} = \tp{\U_1^{*(t)},\hdots,\U_L^{*(t)}, \V_1^{*(t)},\hdots,\V_M^{*(t)}},\] where $\{\U_1^{*(t)},\hdots,\U_L^{*(t)}, \V_1^{*(t)},\hdots,\V_M^{*(t)}\}$ satisfy given restrictions for identifiability.  

For $\tilde{N}$ out-of-sample observations with predictor array $\XX_{\new}: \tilde{N} \times P_1 \ctimes P_L$, the point prediction for the responses is
\begin{align}
\hat{\YY}_{\new} = \tprod{\XX_{\new}}{\hat{\BB}}{L}     \label{pointPred} 
\end{align}
where $\BB$ is given by (\ref{ridgeEq}).  Uncertainty in this prediction can be assessed using samples from the posterior predictive distribution of $\YY_{\new}$: 
\begin{align}
\YY_{\new}^{(t)} = \tprod{\XX_{\new}}{\BB^{(t)}}{L} +\EE_{\new}^{(t)},     \label{postPred} 
\end{align}
where $\EE_{\new}^{(t)}$ is generated with independent N$(0,\sigma^{2 (t)})$ entries.


\section{Simulation study}
\label{sims}

\subsection{Approach}
\label{sims_approach}
We conduct a simulation study to predict a three-way array $\YY$ from another three-way array $\XX$ under various conditions. We implement a fully crossed factorial simulation design with the following manipulated conditions: 
\begin{itemize}
\item Rank $R = 0,1,2,3,4$ or $5$ (6 levels)
\item Sample size $N = 30$ or $120$ (2 levels)
\item Signal-to-noise ratio $\mbox{SNR} = 1$ or $5$ (2 levels).	
\end{itemize}
For each of the $24$ scenarios, we simulate data as follows:
\begin{enumerate}
\item Generate $\XX: N \times P_1 \times P_2$ with independent $N(0,1)$ entries.
\item Generate $\U_l: P_l \times R$ for $l=1,\hdots,L$ and $\V_m: Q_m \times R$ for $m=1,\hdots,M$, each with independent $N(0,1)$ entries.
\item Generate error $\EE: N \times Q_1 \times Q_2$ with independent $N(0,1)$ entries.
\item Set $\YY = \tprod{\XX}{\BB}{L} + \EE$, where 
\[\BB = c \tp{\U_1,\hdots,\U_L, \V_1,\hdots,\V_M}\]
and $c$ is the scalar giving 
\[\frac{||\tprod{\XX}{\BB}{L}||^2_F}{||\EE||^2_F} = \mbox{SNR}.\]
\end{enumerate}
 
We fix the dimensions $p_1=15, p_2=20, q_1=5, q_2=10$, and generate $10$ replicated datasets as above for each of the $24$ scenarios, yielding $240$ simulated datasets.  For each simulated dataset, we estimate $\BB$ as in Section~\ref{opt} under each combination of the following parameters: 
\begin{itemize}
\item Assumed rank $\hat{R}=1,2,3,4$ or $5$ (5 levels)
\item Regularization term $\lambda = 0, 0.5, 1, 5$ or $50$ (5 levels).
\end{itemize}

For each of the $240$ simulated datasets and $5 \cdot 5 = 25$ estimation procedures, we compute the relative out-of-sample prediction error   of the resulting coefficient estimate $\hat{\BB}$.  This is done empirically by generating a new dataset with $\tilde{N}=500$ observations:  
\[\YY_{\new} = \tprod{\XX_{\new}}{\BB}{L} + \EE_{\new}\]
where $\XX_{\new}$ and $\EE_{\new}$ have independent $N(0,1)$ entries.  The relative prediction error (RPE) for these test observations is 
\begin{align}
\mbox{RPE} = \frac{||\YY_{\new}-\tprod{\XX_{\new}}{\hat{\BB}}{L} ||^2_F}{||\YY_{\new}||^2_F}. \label{RPE}
\end{align}

Symmetric 95\% credible intervals are created for each value of $\YY_{\new}$ using $T=1000$ outcome arrays simulated from the posterior (\ref{postPred}).    
 
\subsection{Results}
\label{sims_results}

First we consider the results for those cases with no signal, $R=0$, where the oracle RPE is $1$.  The marginal mean RPE across the levels of $N$, $\lambda$, and $\hat{R}$ are shown in Table~\ref{tab:NoSig}.  Overall, simulations with a higher training sample size $N$ resulted in lower RPE, estimation with higher regularization parameter $\lambda$ resulted in lower RPE, and estimation with higher assumed rank resulted in higher RPE.  These results are not surprising, as a lower sample size, higher assumed rank and less regularization all encourage over-fitting. 

\begin{table}
\caption{\label{tab:NoSig} Marginal mean RPE for no signal, $R=0$.}
\centering
\begin{tabular}{l|c c c c c c c c c c}
\hline 
\textbf{Training samples} & & \multicolumn{3}{c}{$N=30$} & \multicolumn{3}{c}{$N=120$} &  & & \\
& & \multicolumn{3}{c}{1.48} &   \multicolumn{3}{c}{1.08} &  & &\\ \hline
\textbf{Regularization} & \multicolumn{2}{c}{$\lambda=0$} & \multicolumn{2}{c}{$\lambda=0.5$} &  \multicolumn{2}{c}{$\lambda=1$} & \multicolumn{2}{c}{$\lambda=5$} & \multicolumn{2}{c}{$\lambda=50$}\\
& \multicolumn{2}{c}{2.00} & \multicolumn{2}{c}{1.14} &  \multicolumn{2}{c}{1.13} & \multicolumn{2}{c}{1.08} & \multicolumn{2}{c}{1.02} \\ \hline 
\textbf{Assumed rank} & \multicolumn{2}{c}{$\hat{R}=1$} & \multicolumn{2}{c}{$\hat{R}=2$} &  \multicolumn{2}{c}{$\hat{R}=3$} & \multicolumn{2}{c}{$\hat{R}=4$} & \multicolumn{2}{c}{$\hat{R}=5$}\\
& \multicolumn{2}{c}{1.04} & \multicolumn{2}{c}{1.12} &  \multicolumn{2}{c}{1.20} & \multicolumn{2}{c}{1.32} & \multicolumn{2}{c}{1.70} \\ \hline 
\end{tabular}
\end{table}

Table~\ref{tab:lambda} shows the effect of the regularization parameter $\lambda$ on the accuracy of the estimated model, in terms of RPE and coverage rates, for different scenarios. As expected, prediction error is generally improved in scenarios with a higher training sample size and higher signal-to-noise ratio. Higher values of $\lambda$ generally improve predictive performance when the sample size and signal-to-noise ratio are small, as these scenarios are prone to over-fitting without regularization. However, large values of $\lambda$ can lead to over-shrinkage of the estimated coefficients and introduce unnecessary bias, especially in scenarios that are less prone to over-fitting. Coverage rates of the 95\% credible intervals are generally appropriate, especially with a higher training sample size.  However, for the scenario with low sample size and high signal ($N=30, \lambda=120$) coverage rates for moderate values of $\lambda$ are poor, as inference is biased toward smaller values of $\BB$.    

\begin{table}
\caption{\label{tab:lambda} The top panel shows mean RPE by regularization for different scenarios using correct assumed ranks.  The bottom panel shows the coverage rate for 95\% credible intervals, and their mean length relative to the standard deviation of $\YY$.}
\centering
\begin{tabular}{l|c c c c c}
\hline 
\textbf{RPE} (std error) & $\mathbf{\lambda=0}$ &  $\mathbf{\lambda=0.5}$ & $\mathbf{\lambda=1}$ &  $\mathbf{\lambda=5}$ & $\mathbf{\lambda=50}$ \\
\hline 
$N=120, \mbox{SNR} = 1$ &  $\mathbf{0.52} \, (0.01)$ & $\mathbf{0.52} \, (0.01)$ & $\mathbf{0.52} \, (0.01)$ & $\mathbf{0.52} \, (0.01)$ &  $\mathbf{0.59} \, (0.01)$ \\
$N=120, \mbox{SNR} = 5$ & $\mathbf{0.04} \, (0.01)$ & $\mathbf{0.04} \, (0.01)$ &  $\mathbf{0.04} \, (0.01)$ &  $\mathbf{0.05} \, (0.01)$ & $\mathbf{0.20} \, (0.01)$\\
$N=30, \mbox{SNR} = 1$ & $\mathbf{1.90} \, (0.15)$ & $\mathbf{1.07} \, (0.04)$ & $\mathbf{1.03} \,(0.04)$ & $\mathbf{0.92} \,(0.02)$ &  $\mathbf{0.91} \, (0.01)$ \\
$N=30, \mbox{SNR} = 5$ & $\mathbf{1.64}\, (0.12)$ & $\mathbf{0.74} \, (0.05)$ & $\mathbf{0.70} \, (0.04)$ & $\mathbf{0.63} \, (0.02)$ & $\mathbf{0.77} \, (0.01)$ \\
\hline
\textbf{Coverage} (length) & $\mathbf{\lambda=0}$ &  $\mathbf{\lambda=0.5}$ & $\mathbf{\lambda=1}$ &  $\mathbf{\lambda=5}$ & $\mathbf{\lambda=50}$ \\
\hline 
$N=120, \mbox{SNR} = 1$ &  $\mathbf{0.95}\, (2.79)$ & $\mathbf{0.95} \, (2.79)$ & $\mathbf{0.95} \, (2.79)$ & $\mathbf{0.95} \, (2.79)$ &  $\mathbf{0.94} \, (2.90)$ \\
$N=120, \mbox{SNR} = 5$ & $\mathbf{0.95} \, (0.77)$ & $\mathbf{0.95} \, (0.77)$ &  $\mathbf{0.95} \, (0.77)$ &  $\mathbf{0.94} \, (0.80)$ & $\mathbf{0.91} \, (1.43)$\\
$N=30, \mbox{SNR} = 1$ & $\mathbf{0.98} \, (4.75)$ & $\mathbf{0.95}\, (3.56)$ & $\mathbf{0.94} \, (3.36)$ & $\mathbf{0.91} \, (3.05)$ &  $\mathbf{0.91} \, (3.18)$ \\
$N=30, \mbox{SNR} = 5$ & $\mathbf{0.91} \, (3.10)$ & $\mathbf{0.68} \, (1.11)$ & $\mathbf{0.65} \, (1.04)$ & $\mathbf{0.68} \, (1.10)$ & $\mathbf{0.84} \, (2.10)$ \\
\end{tabular}
\end{table}

Table~\ref{tab:ranks} illustrates the effects of rank misspecification on performance, under the scenario with $N=120$, SNR$=1$ and no regularization ($\lambda=0$).  For each possible value of the true rank $R=1,\hdots,5$, the RPE is minimized when the assumed rank is equal to the true rank.  Predictive performance is generally more robust to assuming a rank higher than the true rank than it is to assuming a rank lower than the true rank.  

\begin{table}[H]
\caption{\label{tab:ranks} Mean RPE by assumed rank for different true ranks for $N=120$, S2N$=1$, and $\lambda=0$}
\centering
\begin{tabular}{l|c c c c c}
\hline 
 & $\hat{R}=1$ & $\hat{R}=2$ & $\hat{R}=3$ & $\hat{R}=4$ & $\hat{R}=5$ \\
\hline 
$R=1$ & $\mathbf{0.50}$ & $0.54$ & $0.55$ & $0.56$ & $0.57$\\
$R=2$ & $0.64$ & $\mathbf{0.50}$ & $0.53$ & $0.53$ & $0.54$ \\
$R=3$ &  $0.71$ & $0.58$ & $\mathbf{0.53}$ & $0.55$ &  $0.57$ \\
$R=4$ & $0.78$ & $0.67$ &  $0.57$ &  $\mathbf{0.51}$ & $0.53$\\
$R=5$ & $0.81$ & $0.68$ &  $0.61$ &  $0.55$ & $\mathbf{0.53}$\\
\end{tabular}
\end{table}

See Appendix~\ref{corSim} for additional simulation results when the predictors $\XX$ or response $\YY$ are correlated.

\section{Application}
\label{app}
We use the tensor-on-tensor regression model to predict attributes from facial images, using the Labeled Faces in the Wild database \citep{learned2016labeled}. The database includes over $13000$ publicly available images taken from the internet, where each image includes the face of an individual.  Each image is labeled only with the name of the individual depicted, often a celebrity, and there are multiple images for each individual. The images are unposed and exhibit wide variation in lighting, image quality, angle, etc.\ (hence ``in the wild").

 Low-rank matrix factorization approaches are commonly used to analyze facial image data, particularly in the context of facial recognition \citep{sirovich1987low,turk1991eigenfaces,vasilescu2002multilinear,kim2007color}.   Although facial images are not obviously multi-linear, the use of multi-way factorization techniques has been shown to convey advantages over simply vectorizing images (e.g., from a $P_1 \times P_2$ array of pixels to a vector of length $P_1P_2$) \citep{vasilescu2002multilinear}.  \citet{kim2007color} show that treating color as another mode within a tensor factorization framework can improve facial recognition tasks with different lighting.   Moreover, the CP factorization has been shown to be much more efficient as a dimension reduction tool for facial images than PCA, and marginally more efficient than the Tucker and other multiway factorization techniques \citep{lockComment}.

\cite{kumar2009attribute} developed an attribute classifier, which gives describable attributes for a given facial image.  These attributes include characteristics that describe the individual (e.g., gender, race, age), that describe their expression (e.g., smiling, frowning, eyes open), and that describe their accessories (e.g., glasses, make-up, jewelry).  These attribute were determined on the Faces in the Wild dataset, as well as other facial image databases.  In total $72$ attributes are measured for each image.  The attributes are measured on a continuous scale;  for example, for the smiling attribute, higher values correspond to a more obvious smile and lower values correspond to no smile. 

Our goal is to create an algorithm to predict the $72$ describable and correlated attributes from a given image that contains a face.  First, the images are \emph{frontalized} as described in \cite{hassner2015effective}.  In this process the unconstrained images are rotated, scaled, and cropped so that all faces appear forward-facing and the image shows only the face.  After this step images are aligned over the coordinates, in that we expect the nose, mouth and other facial features to be in approximately the same location.  Each frontalized image is $90 \times 90$ pixels, and each pixel gives the intensity for colors red, green and blue,  resulting in a multiway array of dimensions $90 \times 90 \times 3$.  We center the array by subtracting the ``mean face" from each image, i.e., we center each pixel triplet ($x \times y \times$ color) to have mean $0$ over the collection of frontalized images.     We standardize the facial attribute data by converting the measurements to z-scores, wherein each attribute has mean zero and standard deviation $1$ over the collection of faces.
 
To train the predictive model we use a use a random sample of $1000$ images from unique individuals.  Thus the predictor array of images $\XX$ is of dimension $1000 \times 90 \times 90 \times 3$, and the outcome array of attributes $\YY$ is of dimension $1000 \times 72$.  Another set of $1000$ images from unique individuals are used as a validation set, $\XX_{\new}: 1000 \times 90 \times 90 \times 3$ and $\YY_{\new}: 1000 \times 72$.

\begin{figure}[H]
\centering
\includegraphics[width=0.9\textwidth]{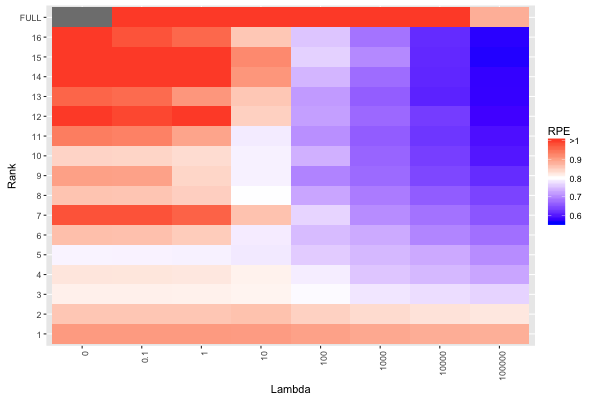}
\caption{\label{fig:heatmap} Relative prediction error for characteristics of out-of-sample images for different parameter choices.  The top row (full rank) gives the results under separate ridge regression models for each outcome without rank restriction.}
\end{figure}

We run the optimization algorithm in Section~\ref{opt} to estimate the coefficient array $\BB: 90 \times 90 \times 3 \times 72$ under various values for the rank $R$ and regularization parameter $\lambda$.  We consider all combinations of the values $\lambda=\{0,0.1,1,10,100,1000,10^4,10^5\}$ and $R=\{1,2,\hdots,16\}$.  We also consider the full rank model that ignores multi-way structure, where the coefficients are given by separate ridge regressions for each of the $72$ outcomes on the $90 \cdot 90 \cdot 3 = 24300$ predictors.  For each estimate we compute the relative prediction error (RPE) for the test set (see (\ref{RPE})).  The resulting RPE values over the different estimation schemes are shown in Figure~\ref{fig:heatmap}.  The minimum RPE achieved was $0.568$,  for $R=15$ and $\lambda=10^5$.  The performance of models with no regularization ($\lambda=0$), or without rank restriction (rank=$\mbox{FULL}$), were much worse in comparison.  This illustrates the benefits of simultaneous rank restriction and ridge regularization for high-dimensional multi-way prediction problems.

\begin{figure}[H]
\centering
\includegraphics[width=0.75\textwidth]{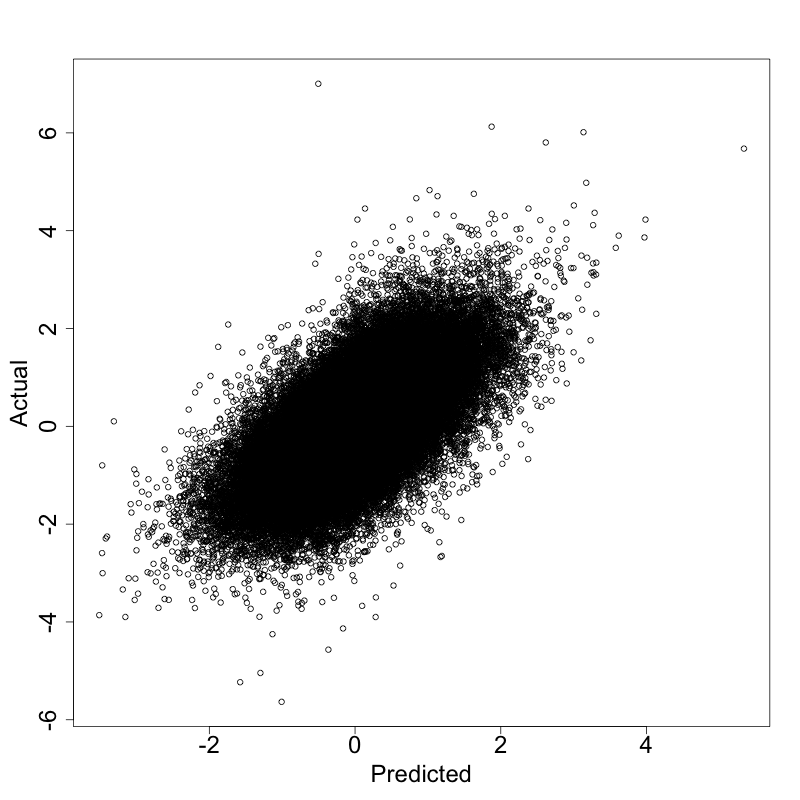}
\caption{\label{fig:scatter} Actual vs. predicted values for $1000$ test images across $72$ characteristics.}
\end{figure}

In what follows we use $R=15$ and $\lambda=10^5$.  Figure~\ref{fig:scatter} shows the predicted values vs.\ the given values, for the test data, over all $72$ characteristics.  The plot shows substantial residual variation but a clear trend, with correlation $r=0.662$.  

\begin{figure}[!h]
\centering
\includegraphics[width=0.49\textwidth]{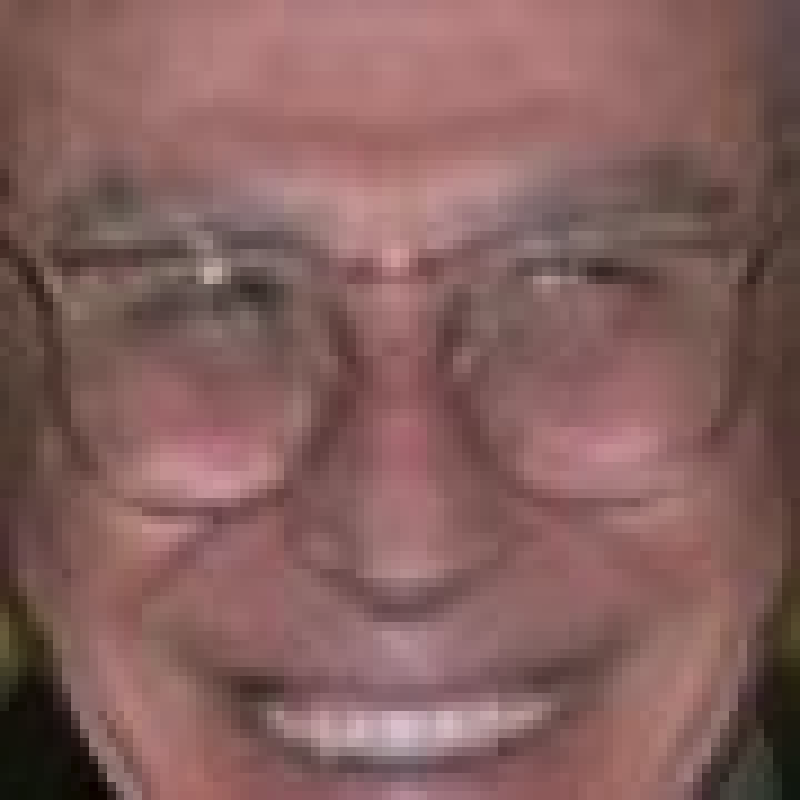}
\includegraphics[width=0.49\textwidth,trim={1.3cm 0 0 0},clip]{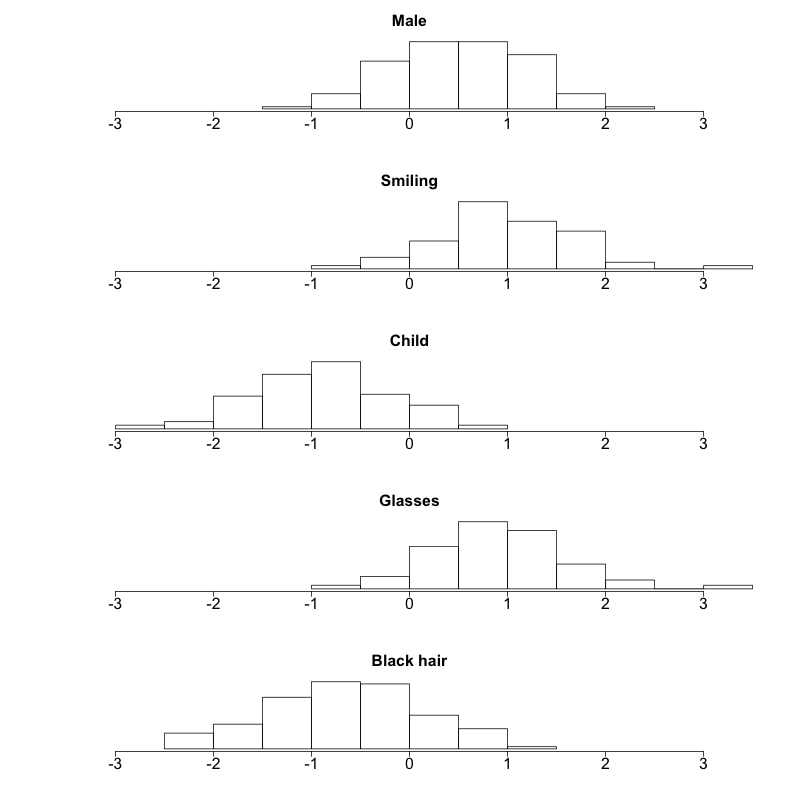}
\caption{\label{fig:postHists} Example test image (left), and its posterior samples for $5$ select characteristics (right).}
\end{figure}

To assess predictive uncertainty we generate $5000$ posterior samples as in Section~\ref{inf}, yielding samples from the posterior predictive distribution of the $72$ characteristics for each of the $1000$ test images.  Symmetric credible intervals were computed for each characteristic of each image. The empirical coverage rates for the given values were $0.934$ for $95\%$ credible intervals and $0.887$ for $90\%$ credible intervals.  The full posterior distributions for a small number of select characteristics, for a single test image, are shown in Figure~\ref{fig:postHists} as an illustration of the results.           

\section{Discussion}
\label{discussion}

In this article we have proposed a general framework for predicting one multiway array from another using the contracted tensor product.  The simulation studies and facial image application illustrate the advantages of CP-rank regularization and ridge regularization in this framework.  These two parameters define a broad class of models that are appropriate for a wide variety of scenarios. The CP assumption accounts for multi-way dependence in both the predictor and outcome array, and the ridge penalty accounts for auxiliary high-dimensionality and multi-collinearity of the predictors.    However, several alternative regularization strategies are possible.  The coefficient array can be restricted to have a more general Tucker structure (as in \citet{li2013}), rather than a CP structure. A broad family of separable penalty functions, such as the separable $L_2$ penalty in (\ref{ridgeEqSep}), are straightforward to impose within the general framework using an alternating estimation scheme similar to that described in \citet{zhou2013tensor}.  In particular, a separable $L_1$ penalty has advantages when a solution that includes sparse subregions of each mode is desired. The alternating estimation scheme described herein for the non-separable $L_2$ penalty is not easily extended to alternative non-separable penalty functions.    

We have described a simple Gaussian likelihood and a prior distribution for $\BB$ that are motivated by the least-squares objective with non-separable $L_2$ penalty.  The resulting probability model involves many simplifying assumptions, which may be over-simplified for some situations.  In particular, the assumption of independent and homoscadastic error in the outcome array can be inappropriate for applications with auxiliary structure in the outcome.  The array normal distribution \citep{akdemir2011array,hoff2011separable} allows for multiway dependence and can be used as a more flexible model for the error covariance.  Alternatively, envelope methods \citep{cook2015foundations} rely on a general technique to account for and ignore immaterial structure in the response and/or predictors of a predictive model.  A tensor envelope is defined in \citet{li2016parsimonious}, and its use in the tensor-on-tensor regression framework is an interesting direction for future work. 

The approach to inference used herein is ``semi-Bayesian", in that the prior is limited to facilitate inference under the given penalized least-squares objective and is not intended to be subjective.  Fully Bayesian approaches, such as using a prior for both the rank of the coefficient array and the shrinkage parameter, are another interesting direction of future work.

\bigskip
\begin{center}
{\large\bf SUPPLEMENTARY MATERIAL}
\end{center}

\begin{description}

\item[MultiwayRegression:] R package MultiwayRegression, containing documented code for all methods described in this article.  (GNU zipped tar file)

\end{description}

\appendix

\section{Consistency}
\label{consistency}

Here we establish the consistency of the minimizer of the objective (\ref{ridgeEq}), for fixed dimension as $N \rightarrow \infty$, under general conditions.  

\begin{theorem}
 Assume model (\ref{arrayEq}) holds for $\BB=\BB_0$, where 
 \begin{enumerate}
 \item For each response index $(q_1,\hdots,q_M)$, the errors $\EE[n,q_1,...,q_M]$ are independent and identically distributed (iid) for $n=1,\hdots,N$, with mean $0$ and finite second moment. 
\item For each predictor index $(p_1,\hdots,p_L)$, $\XX[n,p_1,\hdots,p_L]$ are iid for $n=1,\hdots,N$ from a bounded distribution.   
\item $\BB_0$ has a rank $R_0$ factorization (\ref{Bfac}), where $\theta_0=\{\U_1, \hdots, \U_L, \V_1, \hdots, \V_M\}$ is in the interior of a compact space $\Theta$ and is identifiable under the restrictions of Section~\ref{ident}.  
 \end{enumerate}
For $R=R_0$ and fixed ridge penalty $\lambda\geq0$, the minimizer of the objective \ref{ridgeEq}, $\hat{\beta}_N$, converges to $\BB_0$ in probability as $N\rightarrow \infty$.  Moreover, under the restrictions of Section~\ref{ident} the factorization parameters $\hat{\theta}_N$ converge to $\theta_0$ in probability as $N \rightarrow \infty$.  
\end{theorem}

For Theorem 1 we require that the observations are iid, but within an observation the elements of the error array $\EE$ or predictor array $\XX$ may be correlated or from different distributions.  Also, note that the correct rank is assumed for the estimator, but the result holds for any fixed penalty $\lambda \geq 0$.   The requirements that the predictors $\XX$ are bounded and that $\Theta$ is compact are similar to those used to show the constancy of tensor regression under a normal likelihood model in \citet{zhou2013tensor}.  These requirements facilitate the use of Glivenko-Cantelli theory with a classical result on the asymptotic consistency of M-estimators \citep{van2000asymptotic}; the proof is given below.        

\begin{proof}
Let $\BB(\theta)$ be the coefficient array resulting from $\theta=\{\U_1, \hdots, \U_L, \V_1, \hdots, \V_M\}$: 
\[\BB(\theta) = \tp{\U_1,\hdots,\U_L,\V_1,\hdots,\V_M}.\]
Let $M(\theta)$ be the expected squared error  for a single observation: 
\[M(\theta) = E\left(||\YY_n - \tprod{\XX_n}{\BB(\theta)}\, ||_F^2 \right),\]
which exists for all $\theta \in \Theta$ because the entries of $\EE$ are assumed to have finite second moment.   
Let $M_N^\lambda(\theta)$ be the penalized empirical squared error loss (\ref{ridgeEq}) divided by $N$:  
\[M_N^\lambda(\theta) = \frac{1}{N} \left(\sum_{n=1}^N ||\YY_n - \tprod{\XX_n}{\BB(\theta)}||_F^2\right) + \frac{\lambda}{N} ||\BB(\theta)||^2_F.\]
From Theorem 5.7 of \cite{van2000asymptotic}, the following three properties imply  $\hat{\theta}$ is a consistent estimator of $\theta_0$:
\begin{enumerate}
\item $\underset{\theta: d(\theta,\theta_0) \geq \epsilon}{\inf} M(\theta)>M(\theta_0)$ for any $\epsilon>0$, 
\item $M_N^\lambda(\hat{\theta}_n) \leq M_N^\lambda(\theta_0)-O_P(1)$, where $O_P(1)$ defines a stochastically bounded sequence of random variables, and
\item $\underset{\theta \in \Theta}{\sup} |M_N^\lambda(\theta)-M(\theta)| \overset{P}{\rightarrow} 0$.  
\end{enumerate}
Because $E(\YY) = \tprod{\XX}{\BB(\theta_0)}{L}$,  the coefficient array $\BB(\theta_0)$  minimizes the expected squared error.  Property 1\ then follows from the identifiability of $\theta_0$.  For any $\theta \in \Theta$, $M_N^\lambda(\theta) \rightarrow M(\theta)$ almost surely by the strong law of large numbers and the fact 
\begin{align} \underset{N\rightarrow \infty}{\lim} \,  \underset{\theta \in \Theta}{\sup} \, \frac{\lambda}{N}  ||\BB(\theta)||^2_F = 0.\label{LambdaLim} \end{align}  Also, $M(\theta)$ is necessarily bounded over the compact space $\Theta$.  Thus, both $M_N(\hat{\theta}_n)$ and        
 $M_n(\theta_0)$  are stochastically bounded, and property 2\ follows.  
For property 3\, it suffices to show uniform convergence of the unpenalized squared error $M_N^0 (\theta)$, by
\[\underset{\theta \in \Theta}{\sup} |M_N^\lambda(\theta)-M(\theta)| \leq \underset{\theta \in \Theta}{\sup} |M_N^0(\theta)-M(\theta)|+ \underset{\theta \in \Theta}{\sup}\frac{\lambda}{N} ||\BB(\theta)||^2_F \]
and (\ref{LambdaLim}).  The uniform convergence of $M_N^0 (\theta)$ can be verified by Glivenko-Cantelli theory. Define \[m_\theta(\XX_n,\YY_n) =  ||\YY_n - \tprod{\XX_n}{\BB(\theta)} \,||_F^2.\] The class $\{m_\theta: \theta \in \Theta\}$ is Glivenko-Cantelli, because $\Theta$ is compact and  $m_\theta(\XX_n,\YY_n)$ is continuous as a function of $\theta$ and bounded on $\Theta$ for any $(\XX_n,\YY_n)$.  Thus, property 3\ holds and $\hat{\theta}$ is a consistent estimator of $\theta$.  

By the continuous mapping theorem,  $\BB(\hat{\theta})$ is also consistent estimator of the true coefficient array $\BB(\theta)$. 
\end{proof}

\section{Posterior derivations} 
\label{appPost}
Here we derive the full conditional distributions of the factorization components for \\$\BB=\tp{\U_1,\hdots,\U_L,\V_1,\hdots,\V_M}$, used in Section~\ref{inf}.  

	First, we consider the \emph{a priori} conditional distributions that are implied by the spherical Gaussian prior for $\BB$ (\ref{Bprior}).   Here we derive the prior conditional for $\U_1$, $\pr(\U_1 \mid \U_2,\hdots,\U_L, \V_1,\hdots,\V_M)$;  the prior conditionals for $\U_2,\hdots,\U_L, \V_1,\hdots,\V_M$ are analogous, because the prior for $\BB$ is permutation invariant over its $L+M$ modes. Let  $\b_r^{(1)}$ give the vectorized form of the CP factorization without $\U_1$,
	\[\b_r^{(1)} = \vect \left(\u_{2r} \circ \cdots \circ \u_{Lr} \circ \v_{1r} \circ \cdots \circ \v_{Mr}\right),\]
  and define the matrix $\B^{(1)}: Q \prod_{l=2}^L P_l \times R$ by $\B^{(1)} = \left[\b_1^{(1)} \, \hdots \, \b_R^{(1)} \right]$.  Then 
\[ \vect \left(\U_1 \B^{(1)^T} \right) = \vect (\BB) \sim N\left(\mathbf{0}, \frac{\sigma^2}{\lambda} \I_{PQ \times PQ}\right), \]   
and it follows that 
\begin{align*}
\pr(\vect(\U_1) \mid \U_2,\hdots,\U_L, \V_1,\hdots,\V_M) = N \left(\mathbf{0}, \left(\B^{(1)^T} \B^{(1)}\right)^{-1} \otimes  \frac{\sigma^2}{\lambda}  \I_{P_1 \times P_1} \right ).
\end{align*}

The general model (\ref{arrayEq}) implies
\[\C \vect(\U_1) + \vect (\EE) = \vect (\YY),  \]
where $\C$ is defined as in (\ref{upU}). If $\EE$ has independent $N(0,\sigma^2)$ entries, a direct application of the Bayesian linear model \citep{lindley1972bayes} gives 
\[\pr(\vect(\U_1) \mid \U_2, \hdots,\U_L, \V_1,\hdots,\V_M, \YY, \XX, \sigma^2) = N(\mmu_1, \SSigma_1)\]
where 
\[\mmu_1 = \left(\C^T\C + \lambda \B^{(1)^T} \B^{(1)} \otimes I_{P_1 \times P_1} \right)^{-1}\C^T \vect(\YY)\]
and 
\[\SSigma_1 = \sigma^2 \left(\C^T\C + \lambda \B^{(1)^T} \B^{(1)} \otimes I_{P_1 \times P_1} \right)^{-1}.\]
Basic tensor algebra shows 
\[\B^{(1)^T} \B^{(1)} = \U_2^T \U_2 \cdot \cdots \cdot \U_L^T \U_L \cdot \V_1^T \V_1 \cdot \cdots \cdot  \V_M^T \V_M.\]
The posterior mean and variance for $\U_2,\hdots,\U_L$ are derived in an analogous way.  

For the $\V_m's$ it suffices to consider $\V_M$, as the posterior derivations for $\V_1,\hdots,\V_{M-1}$ are analogous.  The prior conditional for $\V_M$ is
\begin{align*}
\pr(\vect(\V_M) \mid \U_1,\hdots,\U_L, \V_1,\hdots,\V_{M-1}) = N \left(\mathbf{0}, \left(\B^{(L+M)^T} \B^{(L+M)}\right)^{-1} \otimes  \frac{\sigma^2}{\lambda}  \I_{Q_M \times Q_M} \right )
\end{align*}
 and the general model (\ref{arrayEq}) implies 
 \[ \D \V_M + \E = \Y_m,\]
 where $\D$ and $\Y_M$ are defined as in (\ref{upV}), and $\E$ has independent $N(0,\sigma^2)$ entries.  Separate applications of the Bayesian linear model for each row of $\V_M$ gives 
\[\pr(\vect(\V_M) \mid \U_1, \hdots,\U_L, \V_1,\hdots,\V_{M-1}, \YY, \XX, \sigma^2) = N(\mmu_{L+M}, \SSigma_{L+M})\]
where 
\[\mmu_{L+M} = \vect ((\D^T \D + \lambda \B^{(L+M)^T}\B^{(L+M)})^{-1} \D^T \Y_M^T) \]
and 
\[\SSigma_{L+M} = \sigma^2 \left(\D^T\D + \lambda \B^{(L+M)^T}\B^{(L+M)}\right)^{-1} \otimes \I_{Q_M \times Q_M}.\]
Basic tensor algebra shows 
\[\B^{(L+M)^T} \B^{(L+M)} = \U_1^T \U_1 \cdot \cdots \cdot \U_L^T \U_L \cdot \V_1^T \V_1 \cdot \cdots \cdot  \V_{M-1}^T \V_{M-1}.\]
  
\section{Proof of Proposition 1} 
\label{propproof}

Here we prove the equivalence of separable $L_2$ penalization and nuclear norm penalization stated in Proposition~\ref{prop1}.  The result is shown for predicting a vector from a three-way array, in which $\BB=\U_1 \U_2^T$. Analogous results exist for predicting a matrix from a matrix ($\BB= \U_1 \V_1^T$) and predicting a three-way array from a vector ($\BB= \V_1 \V_2^T$).  

In the solution to
\begin{align}
 \argmin{\rank(\BB)\leq R} ||\mathbf{y}- \tprod{\XX}{\BB}{2}||_F^2 + \lambda \sum_{l=1}^2||\U_l||_F^2 \label{sep2}
 \end{align}
the columns of $\U_1$ and $\U_2$, $\{\u_{11},\hdots, \u_{1R}\}$ and  $\{\u_{21},\hdots, \u_{2R}\}$, must satisfy 
\begin{align}
||\u_{1r}||^2=||\u_{2r}||^2=||\u_{1r} \u_{2r}^T||_F  \, \, \text{ for } \, \, r=1,\hdots,R. \label{equal}	
\end{align}
 Here (\ref{equal}) follows from the general result that for $c>0$,
\[\argmin{} \{(a,b) : ab = c\} \, \, a^2+b^2  = (\sqrt{c},\sqrt{c}),\] 
where $a = ||\u_{1r}||^2$, $b = ||\u_{2r}||^2$, and $c = ||\u_{1r} \u_{2r}^T||_F^2$.  Thus,
\begin{align}
\sum_{l=1}^2||\U_l||_F^2 &= \sum_{l=1}^2 \sum_{r=1}^R ||\u_{lr}||^2 \nonumber \\
&= 2 \sum_{r=1}^R ||\u_{1r} \u_{2r}^T||_F, \label{singval}
\end{align}
Under orthogonality of the columns of $\U_1$ and $\U_2$, the non-zero singular values of $\BB$ are $\{||\u_{1r} \u_{2r}^T||_F\}_{r=1}^R$, and therefore (\ref{singval}) is equal to $2 ||\BB||_*$.  It follows that (\ref{sep2}) is equivalent to 
\[\argmin{\rank(\BB)\leq R} ||\mathbf{y}- \tprod{\XX}{\BB}{2}||_F^2 + 2 \lambda ||\BB||_*.\]     

\section{Correlated data simulation}   
\label{corSim}   
 Here we describe the results of a simulation study analogous to that in Section~\ref{sims}, but with correlation in the predictors $\XX$ or in the response $\YY$.  We simulate Gaussian data with an exponential spatial correlation structure using the R package {\tt fields} \citep{fields}. The entries of $\EE$ are assumed to be on a $Q_1 \times Q_2$ grid ($Q_1=5$, $Q_2=10$) with adjacent entries having distance $1$.  The entries of $\XX$ are assumed to be on a $P_1 \times P_2$ grid ($P_1=15$, $P_2=20$) with adjacent entries having distance $1$. The correlation between adjacent locations is $\rho=0.6$ for each scenario, and the marginal variance of the entries is $1$.  Thus, data are simulated exactly as in Section~\ref{sims_approach}, except for the correlation structure of $\XX$ (step 1.) or $\EE$ (step 3.).
 
The resulting RPE and credible interval coverage rates are shown in Table~\ref{tab:lambdaCorX}, which is analogous to Table~\ref{tab:lambda} for the uncorrelated case.  Interestingly, for penalized estimation and $n=30$ the scenario with correlated $\XX$  gives significantly better performance in terms of RPE than the scenario without correlation.  This was unexpected, but may be because correlation in $\XX$ discourages the algorithm from converging to a local minimum.  For correlated $\EE$ the results are often similar to the uncorrelated scenario but tend toward lower accuracy.  In particular, the credible intervals tend to undercover more than for the uncorrelated scenario, or for the scenario with correlated $\XX$. This is probably because correlation in $\EE$ violates the assumed likelihood model for inference, while correlation in $\XX$ does not.             
   
\begin{table}
\caption{\label{tab:lambdaCorX} Mean RPE by regularization and coverage rate for correlated $\XX$ or correlated $\EE$ using correct assumed ranks.  The coverage rates are for 95\% credible intervals, and their mean length relative to the standard deviation of $\YY$ is shown. }
\centering
\begin{tabular}{l|c c c c c}
\hline 
\multicolumn{6}{c}{Correlated $\XX$} \\
\hline 
\textbf{RPE} (std error) & $\mathbf{\lambda=0}$ &  $\mathbf{\lambda=0.5}$ & $\mathbf{\lambda=1}$ &  $\mathbf{\lambda=5}$ & $\mathbf{\lambda=50}$ \\
\hline 
$N=120, \mbox{SNR} = 1$ &  $\mathbf{0.52} \, (0.01)$ & $\mathbf{0.52} \, (0.01)$ & $\mathbf{0.52} \, (0.01)$ & $\mathbf{0.52} \, (0.01)$ &  $\mathbf{0.58} \, (0.01)$ \\
$N=120, \mbox{SNR} = 5$ & $\mathbf{0.04} \, (0.01)$ & $\mathbf{0.04} \, (0.01)$ &  $\mathbf{0.04} \, (0.01)$ &  $\mathbf{0.05} \, (0.01)$ & $\mathbf{0.19} \, (0.01)$\\
$N=30, \mbox{SNR} = 1$ & $\mathbf{1.62} \, (0.11)$ & $\mathbf{0.87} \, (0.02)$ & $\mathbf{0.83} \,(0.02)$ & $\mathbf{0.79} \,(0.02)$ &  $\mathbf{0.80} \, (0.01)$ \\
$N=30, \mbox{SNR} = 5$ & $\mathbf{1.70}\, (0.17)$ & $\mathbf{0.43} \, (0.03)$ & $\mathbf{0.43} \, (0.03)$ & $\mathbf{0.45} \, (0.02)$ & $\mathbf{0.58} \, (0.01)$ \\
\hline
\textbf{Coverage} (length) & $\mathbf{\lambda=0}$ &  $\mathbf{\lambda=0.5}$ & $\mathbf{\lambda=1}$ &  $\mathbf{\lambda=5}$ & $\mathbf{\lambda=50}$ \\
\hline 
$N=120, \mbox{SNR} = 1$ &  $\mathbf{0.95}\, (2.80)$ & $\mathbf{0.95} \, (2.80)$ & $\mathbf{0.95} \, (2.80)$ & $\mathbf{0.95} \, (2.80)$ &  $\mathbf{0.94} \, (2.92)$ \\
$N=120, \mbox{SNR} = 5$ & $\mathbf{0.95} \, (0.78)$ & $\mathbf{0.95} \, (0.78)$ &  $\mathbf{0.95} \, (0.78)$ &  $\mathbf{0.94} \, (0.83)$ & $\mathbf{0.92} \, (1.42)$\\
$N=30, \mbox{SNR} = 1$ & $\mathbf{0.98} \, (4.75)$ & $\mathbf{0.95}\, (3.56)$ & $\mathbf{0.94} \, (3.36)$ & $\mathbf{0.92} \, (3.05)$ &  $\mathbf{0.92} \, (3.18)$ \\
$N=30, \mbox{SNR} = 5$ & $\mathbf{0.93} \, (3.10)$ & $\mathbf{0.76} \, (1.11)$ & $\mathbf{0.74} \, (1.04)$ & $\mathbf{0.74} \, (1.10)$ & $\mathbf{0.84} \, (2.10)$ \\
\hline 
\multicolumn{6}{c}{Correlated $\EE$} \\
\hline
\textbf{RPE} (std error) & $\mathbf{\lambda=0}$ &  $\mathbf{\lambda=0.5}$ & $\mathbf{\lambda=1}$ &  $\mathbf{\lambda=5}$ & $\mathbf{\lambda=50}$ \\
\hline 
$N=120, \mbox{SNR} = 1$ &  $\mathbf{0.59} \, (0.01)$ & $\mathbf{0.59} \, (0.01)$ & $\mathbf{0.59} \, (0.01)$ & $\mathbf{0.58} \, (0.01)$ &  $\mathbf{0.60} \, (0.01)$ \\
$N=120, \mbox{SNR} = 5$ & $\mathbf{0.04} \, (0.01)$ & $\mathbf{0.04} \, (0.01)$ &  $\mathbf{0.04} \, (0.01)$ &  $\mathbf{0.04} \, (0.01)$ & $\mathbf{0.19} \, (0.01)$\\
$N=30, \mbox{SNR} = 1$ & $\mathbf{1.86} \, (0.13)$ & $\mathbf{1.10} \, (0.02)$ & $\mathbf{1.04} \,(0.02)$ & $\mathbf{0.93} \,(0.01)$ &  $\mathbf{0.91} \, (0.01)$ \\
$N=30, \mbox{SNR} = 5$ & $\mathbf{2.03}\, (0.22)$ & $\mathbf{0.72} \, (0.03)$ & $\mathbf{0.68} \, (0.03)$ & $\mathbf{0.63} \, (0.02)$ & $\mathbf{0.77} \, (0.01)$ \\
\hline 
\textbf{Coverage} (length) & $\mathbf{\lambda=0}$ &  $\mathbf{\lambda=0.5}$ & $\mathbf{\lambda=1}$ &  $\mathbf{\lambda=5}$ & $\mathbf{\lambda=50}$ \\
\hline 
$N=120, \mbox{SNR} = 1$ &  $\mathbf{0.93}\, (2.75)$ & $\mathbf{0.93} \, (2.75)$ & $\mathbf{0.93} \, (2.75)$ & $\mathbf{0.93} \, (2.75)$ &  $\mathbf{0.93} \, (2.88)$ \\
$N=120, \mbox{SNR} = 5$ & $\mathbf{0.95} \, (0.77)$ & $\mathbf{0.95} \, (0.77)$ &  $\mathbf{0.95} \, (0.77)$ &  $\mathbf{0.94} \, (0.79)$ & $\mathbf{0.91} \, (1.41)$\\
$N=30, \mbox{SNR} = 1$ & $\mathbf{0.98} \, (4.89)$ & $\mathbf{0.94}\, (3.51)$ & $\mathbf{0.93} \, (3.28)$ & $\mathbf{0.89} \, (2.92)$ &  $\mathbf{0.91} \, (3.12)$ \\
$N=30, \mbox{SNR} = 5$ & $\mathbf{0.91} \, (2.98)$ & $\mathbf{0.68} \, (1.10)$ & $\mathbf{0.66} \, (1.01)$ & $\mathbf{0.67} \, (1.05)$ & $\mathbf{0.84} \, (2.09)$ \\
\end{tabular}
\end{table}

\bibliographystyle{apa}

\bibliography{tensor-on-tensor.bib}
\end{document}